\documentclass[12pt,a4paper,epsf,epsfig]{article}
\usepackage{epsfig}
\usepackage{feynmf}
\usepackage[english]{babel}
\begin{document}
\large
\date{}
\title {\bf CONTRIBUTION OF SUPERSTRING $Z^{\prime}$ BOSON INTO THE
POLARIZATION EFFECTS IN PROTON PROTON COLLISIONS}

\author {R.~Kh.~Muradov$^{1}$ \,\,\,  and \,\,\,
A.~I.~Ahmadov$^{2}$ \footnote {E-mail: ahmadovazar@yahoo.com}}
\maketitle
\begin{center}
$^{1}$ {\small Department of Theoretical Physics, Baku State
University,} \\
$^{2}$ {\small High Energy Physics Lab., Faculty of Physics, Baku
State University, \\ Z.Khalilov st. 23,
AZ-1148, Baku, Azerbaijan} \\
\end{center}
\begin{abstract}
In this work we investigate the single- and the double-spin
asymmetries at the collisions of polarized protons $pp \to
(\gamma^*, Z^0, Z') +X$ within the scope of QCD, electroweak
interaction and superstring $E_6$ theory. The helicity amplitude
method is used. Analytical expressions for the single- and the
double-spin asymmetries are obtained and their dependence from the
transverse momentum of lepton pair is investigated at the three
different values of invariant masses of lepton pair. Pure contribution
coming from superstring $Z'$ boson into the single- and double- spin asymmetries
has been extracted. The obtained results allow to investigate the spin structure of the proton.
\end{abstract}



\section{\bf Introduction}

With the advent of the RHIC at BNL we have a new facility to study
the spin structure of the proton,(for a review on the potential of
RHIC [1]), which supplements the existing polarized lepton-hadron
machines. Polarized proton-proton collisions with a very high
luminosity and a maximum centre of mass energy 500 GeV will
provide us with many more details about spin distributions than
possible with the existing lepton-hadron machines, which give very
little information about the polarized gluon and sea-quark parton
densities. The production of lepton pairs in hadron collisions,
the Drell-Yan process [2], is one of the most powerful tools to
probe the structure of hadrons. Its parton model interpretation is
straightforward-the process is induced by the annihilation of a
quark-antiquark pair into a virtual photon which subsequently
decays into a lepton pair. The Drell-Yan process in proton-proton
or proton-nucleus collisions therefore provides a direct probe of
the antiquark densities in protons and nuclei. Since the
appearance of the data obtained in the EMC experiment [3], many
studies have been devoted to the spin structure of the proton. In
the lowest order of perturbative QCD, the spin of the proton can
be represented as the sum of three terms; that is,
$$S_p=S_q+S_g+ \langle L_z \rangle,$$
where $S_q$ and $S_g$ are, quark and gluon contributions to the proton
spin  respectively, and $\langle L_z\rangle $ is the
contribution of the orbital angular momentum of quarks and
gluons.

The experimental results reported in [4-6] suggest that the gluon
spin and orbital interaction contribute significantly to the
proton spin. Naturally, these experimental results require a
theoretical explanation.

Gehrmann [7] calculated $O(\alpha_s)$ correction to $x_F$ and
$y$ distributions of dileptons produced in collisions of
longitudinally polarized hadrons. He also showed that measurement
of the longitudinally polarized cross section for the Drell-Yan
process would make it possible to investigate the distribution of
polarized sea quarks in hadrons.

In [8], the longitudinal-transverse spin asymmetries $(A_{LT})$ in
Drell-Yan processes were calculated in the leading order for
nucleon-nucleon collisions at RHIC energies. It was shown that $A_{LT}$
is much less than the respective transverse-transverse asymmetry $A_{TT}$.

The Drell-Yan process at high transverse momenta of the dilepton
was studied in [9], where the effect of $\gamma$ - $ Z^{0}$
-interference was also taken into consideration. Both single-spin
and double-spin asymmetries were investigated there. It was shown
that the double-spin asymmetry at small invariant masses of the
lepton pair and the single-spin asymmetry at the $Z^{0}$ - peak
becomes significant.

Also, as it was demonstrated in our work [10] single-spin and
double -spin asymmetries in $pp \to (\gamma^*,Z^{0})+X$ processes
induced by collisions of polarized hadrons are analyzed on the
basis of QCD and electroweak interaction. Here it is shown that
the contribution of the $Z^{0}$ boson processes into polarized
particles, in general, and  into the single-spin parity-violating
asymmetries in particular, is significant. The single-spin
asymmetries become large, both in the CLW and in the GRSW models,
as the dilepton invariant mass approaches the $Z^{0}$- boson mass.
Double-spin asymmetry for all the considered $p_T$ is greather
than the single-spin asymmetry for all the values of the dilepton
invariant mass.

Collisions of polarized hadrons are among processes of greater
importance for study of the spin structure of the proton and for
calculation of the distributions of polarized quarks in the proton.

The remainder of this paper is organised as follows:in section 2
we shall give some formulae for the amplitudes and the
differentrial cross section for calculations single-and the
double-spin asymmetries,in section 3 we present the numerical
results and discuss the asymmetries. In Section 4, we draw our
conclusions.

In the present paper we investigate contribution  superstring $Z^{\prime}$
boson to the single-spin and double-spin asymmetries in $pp \to (\gamma^*,Z^{0},Z^{\prime})+X$
processes for studying the spin structure of the proton.

There is currently vigorous research on $E_6$ theory in
elementary particle physics. In the low-energy limit, the $E_6$
group may be split up into subgroups of fifth or sixth ranks [11]:
$$
G_5 = SU_c(3)\times SU_L(2)U_y(1)\times U_{\eta}(1),
$$
$$
G_6 = SU_c(3) \times SU_L(2) \times U_y(1)\times U_{\psi}(1)
\times U_{\chi}(1).
$$

One additional neutral vector field $Z_{\eta}$ arises in group
$G_{5}$, which correspond to $U_{\eta}(1)$ symmetry. In group
$G_6$, we have two additional neutral boson fields $Z_{\psi}$ and
$Z_{\chi}$, which correspond to $U_{\psi}(1)$ and $U_{\chi}(1)$
symmetries respectively. A model having the gauge group $SU_c(3)
\times SU_L(2) \times U_y(1) \times U(1)$ is now considered as the
permissible low-energy limit in superstring theory. An additional
$Z'$ boson arises in this model,
$$
Z'=Z_{\psi} \cos \theta_E + Z_{\chi} \sin \theta_E
$$
in which  $\theta_E$ is the mixing angle ; $\theta_{E}$ is
arbitrary for $G_6$ group, while $\theta_{E} =142.24^{\circ}$ for
$G_5$. The following is the lagrangian for the interaction of the
fundamental fermions with the gauge bosons:
\begin{equation}
L=\frac{e}{2}(J_{\mu}^{\gamma} A_{\mu}+J_{\mu}^{Z^{0}}
{Z_{\mu}^{0}}+J_{\mu}^{Z'} Z_{\mu}'),
\end{equation}
in which
\begin{equation}
J_{\mu}^i=\bar \Psi_f \gamma_{\mu} [g_{L_f}^i (1+\gamma_s) +
g_{R_f} (1- \gamma_s)] \Psi_f,
\end{equation}
and $g_{L_f}^i$ and $g_{R_f}^i$ are chiral coupling constants for
the fermion $f$ with the gauge bosons $i (i=\gamma, Z^{0}, Z')$,
which eigenvalue are equal:
$$
g_{L_f}^\gamma=g_{R_f}^\gamma = Q_f,\,\,\,\,g_{L_f}^{Z^{0}} =
\frac{2}{\sin 2\Theta_W}(I_{3}-Q_f X_W);
$$
$$
g_{R_f}^{Z^{0}}=\frac {2}{\sin 2 \Theta_W} (-Q_f X_W);
$$
$$
g_{L_f}^{Z'}=(5/3)^{0.5}\cdot\frac{1}{\cos\Theta_W}
[Q_\psi(f_L)\cos\theta_E+Q_{\chi}(f_L)\sin\theta_E];
$$
$$
g_{R_f}^{Z'}=(5/3)^{0.5}\cdot\frac{1}{\cos\Theta_W}
[-Q_\psi(f_L)\cos\theta_E+Q_{\chi}(f_L)\sin\theta_E];
$$

Here $X_W=\sin^2\Theta_W$ is the Weinberg parameter, while $Q_f$
and $I_{3f}$ are the generators of $U_{\psi}(1)$ and
$U_{\chi}(1)$ groups, which correspond to charge and third
projection of the isospin for fermion
f:
$$
Q_\psi(f_L)=\frac{1}{\sqrt
{24}},\,\,\,\,Q_\chi(f_L)=\frac{3}{\sqrt {40}}
\,\,\,\,\rm{for}\,\,\,\, f_L=(\widetilde{d},e,\nu_e)_L,
$$
$$
Q_\psi(f_L)=\frac{1}{\sqrt{24}},\,\,\,\,
Q_\chi(f_L)=-\frac{1}{\sqrt{40}} \,\,\,\,\rm{for}\,\,\,\,
f_L=(u,d,\widetilde{u},\widetilde{e})_L.
$$
-generators of groups $U_{\psi}(1)$ and $U_{\chi}$

The mass matrix for $Z^{0}$ and $Z'$ as usual is not diagonalized.
This leads to $Z^{0}-Z'$- mixing
$$
\left(\begin{array}{c}Z_1\\Z_2\end{array}\right)=
\left(\begin{array}{cc}\cos\phi&\sin\phi\\-\sin\phi&\cos\phi\end{array}\right)
\cdot\left(\begin{array}{c}Z^{0}\\Z^{\prime}\end{array}\right)
$$
where $Z_1$ and $Z_2$ are gauge bosons with physical masses $m_{Z_1}$ and
$m_{Z_2}$, $\phi$ is an angle mixing, defined by
$$
tg^2 \phi=(m_{Z^0}^2-m_{Z_1}^2) / (m_{Z_2}^2 - m_{Z^0}^2)
$$
$m_{Z^{0}}$  is a mass of $Z^{0}$ boson in SM.
The expression for Lagrangian interaction fermion with bosons $Z_1$ and
$Z_2$ obtained from (1) has a form
$$
L=\frac{e}{2}(J_\mu^{Z_1} Z_{1\mu}+J_\mu^{Z_2} Z_{2\mu})
$$
where currents $J_\mu^{Z_1}$ and $J_\mu^{Z_2}$ have common form (2),
but the couplings are defined by
$$
g_{L(R)f}^{Z_1}= \cos\phi \cdot g_{L(R)f}^{Z^0} +\sin\phi \cdot
g_{L(R)f}^{Z'}
$$
$$
g_{L(R)f}^{Z_2}=-\sin\phi \cdot g_{L(R)f}^{Z^{0}} + \cos\phi \cdot
g_{L(R)f}^{Z'}
$$
This process is performed in the reference frame comoving with the
center of mass of primary particles. In order to describe
experiments that study the scattering of polarized particles, it
is necessary to specify a helicity basis. Here, we use the method
of helicity amplitudes. It should be noted that, in [7,8], the
asymmetries were studied with allowance for the polarizations of
primary particles. Here, we consider asymmetries, taking into
account the polarizations of all particles that participate in the
reaction under study.

\section{\bf Calculation of asymmetry }

We begin our analysis by introducing the subprocesses
\begin{eqnarray}
g+q \to l^+l^-+q, \nonumber \\
q + \overline {q} \to l^+l^-+g.
\end{eqnarray}
The Feynman diagrams of these subprocess are shown in Fig.1. \\
The matrix elements for subprocesses (3) with allowance for a virtual
photon, $Z^0$ and $Z'$- bosons can be represented as
$$
M_{\gamma} = - i e^2 g_s t_{ik}^a \biggl[\overline {v}(p_4, s_4)
\gamma_{\mu}
u(p_3,s_3)\biggr]{{g_{\mu\nu}}\over{q^2}} \overline {u} (p_2,s_2)
\hat {\varepsilon}{{\hat f_1+m} \over{f_1^2-m^2}} \gamma_\nu
u(p_1,s_1),
$$
$$
M_{Z_{n}}=-{{i g^2 g_s t_{ik}^a} \over{ 4\cos^2
\Theta_W}}\sum_{n=1}^{n_Z}\biggl[\overline {v}(p_4,s_4)
\gamma_{\mu} (L_{e_{n}}P_{L}+R_{e_{n}}P_{R}) u(p_3,s_3)\biggr] \times
$$
$$
\times {D_{Z_{n}}(q^{2})} \overline{u}(p_2,s_2) \hat
{\varepsilon} {{\hat  f_1+m} \over {f_1^2 - m^2}} \gamma_{\mu}
(L_{q_n}P_{L}+R_{q_n}P_{R}) u(p_1,s_1);
$$
$$
M_{\gamma} =-i e^2 g_s t_{ik}^a \biggl[\overline{v}
(p_4,s_4)\gamma_{\mu} u(p_3, s_3) \biggr]{{g_{\mu \nu}}\over
{q^2}}\overline {u}(p_2,s_2)\gamma_{\nu}{{\hat
f_2+m}\over{f_2^2-m^2}} \hat {\varepsilon} u(p_1,s_1),
$$
$$
M_{Z_{n}}=-{{ig^2 g_s t_{ik}^a}\over{4\cos^2\Theta_W}}\sum_{n=1}^{n_Z}\biggl
[\overline {v}(p_4,s_4) \gamma_\mu (L_{e_{n}}P_{L}+R_{e_{n}}P_{R})
u(p_3, s_3) \biggr ] \times
$$
$$
\times {D_{Z_{n}}(q^{2})}u(p_2,s_2)
\gamma_{\mu}(L_{q_{n}}P_{L}+R_{q_{n}}P_{R}) {{\hat
f_2+m}\over{f_2^2-m^2}}\hat \varepsilon u(p_1,s_1);
$$

\begin{equation}
M_{\gamma} =-ie^2 g_s t_{ik}^a \biggl [ \overline {v} (p_4,s_4)
\gamma_{\mu} u(p_3,s_3) \biggr ] {{g_{\mu
\nu}}\over{q^2}}\overline {v}(p_2,s_2) \hat  {\varepsilon} {{\hat
f_3+m}\over{f_3^2-m^2}}\gamma_{\nu} u(p_1,s_1),
\end{equation}
$$
M_{Z_{n}}=-{{i g^2 g_s
t_{ik}^a}\over{4\cos^2\Theta_W}}\sum_{n=1}^{n_Z}\biggl[\overline {v} (p_4,s_4)
\gamma_{\mu} (L_{e_{n}}P_{L}+R_{e_{n}}P_{R})u(p_3,s_3) \biggr] \times
$$
$$
\times {D_{Z_{n}}(q^{2})} \overline {v} (p_2,s_2) \hat
{\varepsilon} {{\hat f_3+m} \over{f_3^2-m^2}} \gamma_{\mu}
(L_{q_{n}}P_{L}+R_{q_{n}}P_{R}) u(p_1,s_1);
$$
$$
M_{\gamma}=-i e^2 g_s t_{ik}^a \biggl [\overline
{v}(p_4,s_4)\gamma_{\mu} u(p_3,s_3) \biggr ]
{{g_{\mu\nu}}\over{q^2}}v(p_2,s_2) \gamma_{\nu} {{\hat
f_4+m}\over{f_4^2-m^2}}
\hat \varepsilon u(p_1,s_1),
$$
$$M_{Z_{n}}=-{{i g^2 g_s t_{ik}^a}\over{4\cos^2\Theta_W}}\sum_{n=1}^{n_Z} \biggl
[\overline {v}
(p_4,s_4) \gamma_{\mu} (L_{e_{n}}P_{L}+R_{e_{n}}P_{R}) u(p_3,s_3)
\biggr ] \times
$$
$$
\times {D_{Z_{n}}(q^{2})} \overline {v} (p_2,s_2) \gamma_{\mu}
(L_{q_{n}}P_{L}+R_{q_{n}}P_{R}) {{\hat f_4+m}\over{f_4^2-m^2}}\hat
\varepsilon u(p_1,s_1).
$$
where $t_{ik}^a={{\lambda_{ik}^a}\over{2}}$  are the Gell-Mann
matrices, $g_s$ is the strong-interaction coupling constant,
$$f_1=p_1 - q, \quad f_2=p_1+p_2, \quad f_3=p_1-q, \quad f_4=p_1-p_5.
$$
The helicity amplitudes are denoted by
$M(\lambda_1,\lambda_2;\lambda_3,\lambda_4,\lambda_5)$ , where
$\lambda_1 $ and $\lambda_2$  are helicities of the initial
partons, $\lambda_3, \lambda_4 $ are the helicities of two
leptons, and $\lambda_5$ is helicity of the final state parton:
\begin{eqnarray}
M(\lambda_1, \lambda_2;\lambda_3,\lambda_4,\lambda_5)= \left \{
\begin{array}{lll}
M(\lambda_1, \lambda_2;\lambda_3,-\lambda_3,\lambda_2) & {\rm for} &
{g+q \to l^+l^-+q,} \\
M(\lambda_1, -\lambda_1; \lambda_3,-\lambda_3,\lambda_5)& {\rm for
} & {q+\overline q \to l^+l^-+g}.
\end{array}
\right.
\end{eqnarray}
Positive- and negative-helicity states are denoted by
$|A_\pm \rangle $, they have the following properties:
$$(1+\gamma_5)|A_\pm \rangle  =0, $$
$$ |A_+ \rangle ^c=-|A_-\rangle ,$$
\begin{equation}
\langle  A_\mp|B_\pm \rangle  =- \langle B_\mp|A_\pm \rangle ,
\end{equation}
$$ \langle  A_+|\gamma_\mu|B_+ \rangle  = \langle B_-|\gamma_{\mu}|A_-
\rangle , $$
Making use of the Fierz identities we obtain
\begin{equation}
\langle A_+|\gamma_{\mu}|B_+ \rangle  \langle C_-|\gamma_{\mu}|D_-
\rangle  = 2 \langle A_+|D_- \rangle \langle C_-|B_+ \rangle,
\end{equation}
$$\langle A_-|B_+ \rangle  \langle  C_-|D_+ \rangle  =\langle A_-|D_+
\rangle  \langle  C_-|B_+ \rangle  + \langle
A_-|C_+ \rangle  \langle  B_-|D_+ \rangle.$$
The spinors
$u_\pm(p), v_\pm (p) $ describing a particle of momentum $p$ and
helicity $\lambda=\pm 1$ satisfy the relations
$$\hat  pu(p)=\hat  p v (p)=\overline {u}(p) \hat  p=\overline {v}(p)
\hat  p,\quad  p^2=0, $$
\begin{equation}
(1 \pm \gamma_5) v_\pm  = (1 \mp \gamma_5)u_\pm=\overline
{u}_{\pm} (1 \pm \gamma_5) =\overline {v}_\pm (1 \mp \gamma_5)=0,
\end{equation}
$$\overline {u}_\pm (p) \gamma_{\mu} u_\pm(p)=\overline {v}_\pm
(p)\gamma_\mu
v_\pm (p)=2p_\mu. $$
Here and below we use the conventional notation
$$u_\pm (p)= v_\mp (p)=|p_\pm \rangle,$$
$$\overline {u}_\pm (p)= \overline {v}_\mp(p)=\langle  p_\pm|, \quad $$
\begin{equation}
\langle p_-|q_+ \rangle = \langle pq\rangle =-\langle qp\rangle ,
\end{equation}
$$\langle q_+|p_-\rangle =\langle pq\rangle ^*=-\langle qp\rangle ^*,
$$
$$|\langle pq\rangle |^2=2p \cdot q. $$
The gluon helicities are defined as follows:
\begin{equation}
\varepsilon_1^\pm =\pm{{\sqrt 2}\over{\langle
p_5^\mp|p_1^\pm\rangle }}\biggl[|p_1^\mp \rangle \langle p_5^\mp|+
|p_5^\pm \rangle \langle  p_1^\pm|\biggr]\,\,\,\,\, {\rm
for}\,\,\,\, g+q \to l^+ l^-+q ,
\end{equation}
\begin{equation}
\varepsilon_5^\pm =\pm{{\sqrt 2}\over{\langle p_1^\mp|p_5^\pm
\rangle }}\biggl [| p_5^\mp \rangle  \langle  p_1^\pm| + |p_1^\pm
\rangle \langle p_5^\pm|\biggr]\,\,\,\,\,{\rm for}\,\,\,\,\,
q+\bar{q} \to l^+l^- +g.
\end{equation}
The Mandelstam invariant variables for the subprocess under
consideration are defined as
\begin{equation}
\hat  s=(p_1+p_2)^2, \quad \hat t=(p_5-p_2)^2=(Q-p_1)^2,\quad \hat
u=(p_5-p_1)^2=(Q-p_2)^2.
\end{equation}
Let us consider the reference frame comoving with the center of mass
of primary particles, where the momenta of primary hadrons are given by
$$P_1={{\sqrt s}\over 2}(1,0,0,1), \quad P_2={{\sqrt s}\over 2
}(1,0,0-1),$$
$$p_1=x_1 P_1, \quad p_2=x_2P_2,$$
\begin{equation}
p_5^\mu=p_1^\mu + p_2^\mu-Q^\mu,
\end{equation}
$$Q^\mu=p_3^\mu +p_4^\mu,$$
$$q^\mu=p_3^\mu - p_4^\mu;$$
$p_5 $  is momentum of the outgoing parton. The momenta of two
leptons and of the final state parton are taken to be [12]
$$p_3^\mu={{1}\over{2}}(E'- q' \cos\alpha, q' \sin\theta - q \sin
\alpha \cos\beta \cos\theta -
E' \cos \alpha \sin \theta, $$
$$-q \sin \alpha \sin \beta, q' \cos\theta -E' \cos\alpha
\cos\theta+q\sin\alpha \cos \beta \sin\theta),$$
\begin{equation}
p_4^\mu={{1}\over{2}}(E' + q' \cos\alpha, q' \sin\theta +q
\sin\alpha \cos\beta \cos\theta +E' \cos\alpha \sin\theta,
\end{equation}
$$q \sin\alpha \sin\beta, q' \cos \theta +E' \cos \alpha \cos \theta -q
\sin\alpha \cos\beta \sin\theta),$$
$$p_5=(q',-q \sin\theta, 0,-q' \cos\theta),$$
where $E'=\frac{\hat s+q^2}{2\hat s}$  and $q'=\frac{\hat s-q^2}{2\hat
s}.$ \\

We now proceed to computing the square of the matrix element taking into
account all helicity states of the particles. \\
(i) The diagrams in Figs.1.a and 1.b yield
$$
|M(++;+-+)|^2= \biggl[\sum_{n=1}^{n_z}2| D_{Z_n}(q^2)|^2\, g_s^2 g^4
R_{q_n}^2
L_{e_n}^2 \\
$$
$$
+\frac{8g_s^2 e^4 e_q^2}{q^4}+ \sum_{\scriptstyle n,n'=1
\atop\scriptstyle n < n'}^{n_z} 4\cdot
Re[D_{Z_n}(q^2)D_{Z_{n'}}^\star(q^2)] g_s^2 g^4 R_{q_n} R_{q_{n'}}
L_{e_n} L_{e_{n'}}+
$$
$$ +\sum_{n=1}^{n_Z}
\frac{[|D_{Z_n}(q^2)|^2 8 g_s^2 g^2 e_q
R_{q_n}L_{e_n}(q^2-m_{Z_n}^2)]}{q^2}\biggr]\times
$$
$$
\times \biggl\{{{2\pi}\over{\hat s}}\hat t^2 +{{\pi}\over{\hat
s}}\hat t \hat u +\pi \hat u +2 \pi Q^2+ {{\pi}\over{\hat s}}\hat
u^2 +{{\pi(Q^2\hat t-\hat s \hat u)}\over{3 \hat s \hat u (\hat
s-Q^2)}}(\hat s \hat t \hat u +Q^2 \hat s \hat t-\hat s^2 \hat u-
\hat s \hat u^2-
$$
\begin{equation}
-Q^2 \hat u \hat t -Q^2\hat t^2)+  {{4\pi} \over{3}} {{Q^2(\hat
t^2+ \hat u \hat t-\hat s \hat t)}\over{(\hat s^2-Q^2)}}\biggr\},
\end{equation}

$|M(++;-++)|^2=|M(++;+-+)|^2,$
$$|M(+-;+--)|^2= \biggl[\sum_{n=1}^{n_z} 2| D_{Z_n}(q^2)|^2 \, g_s^2
g^4 L_{q_n}^2
L_{e_n}^2 + \frac{8g_s^2 e^4 e_q^2}{q^4} + $$
$$+\sum_{\scriptstyle n,n'=1 \atop\scriptstyle n<n'}^{n_z} 4
Re[D_{Z_n}(q^2) D_{Z_{n'}}^\star(q^2)] g_s^2 g^4L_{q_n} L_{q_{n'}}
L_{e_n} L_{e_{n'}} + $$
$$+\sum_{n=1}^{n_Z}\frac{[|D_{Z_n}(q^2)|^2 8 g_s^2 g^2 e_q
L_{q_n}L_{e_n}(q^2-m_{Z_n}^2]}{q^2}\biggr]
$$
\begin{equation}
\times \biggl \{ {{-\pi Q^2}\over{\hat s \hat u}}(\hat t + \hat u
+2 Q^2)^2 - {{\pi Q^2} \over {3 \hat s \hat u (\hat s -Q^2)^2}}
(Q^2 \hat t + Q^2 \hat u - \hat s \hat t - \hat s \hat u)^2
\biggr\}
\end{equation}
$|M(+-;-+-)|^2=|M(+-;+--)|^2$ \,\,\,for \,\,\,$R_e \longleftrightarrow
L_e$. \\
(ii) The  contribution of the diagrams in Figs.1.c and 1.d is
$$
|M(+-;+-+)|^2= \biggl[\sum_{n=1}^{n_z}2|D_{Z_n}(q^2)|^2\,g_s^2 g^4
L_{q_n}^2 L_{e_n}^2+\frac{8g_s^2 e^4 e_q^2}{q^4} +$$
$$
\sum_{\scriptstyle n,n'=1  \atop\scriptstyle n<n'}^{n_z} 4 \cdot
Re[D_{Z_n}(q^2) D_{Z_{n'}}^\star(q^2)] g_s^2 g^4L_{q_n}L_{q_{n'}}
L_{e_n}L_{e_{n'}} +
$$
$$
+\sum_{n=1}^{n_Z} \frac {[|D_{Z_n}(q^2)|^2 8 g_s^2 g^2 e_q L_{q_n}
L_{e_n} (q^2 - m_{Z_n}^2)]} {q^2} \biggr] \times
$$
\begin{equation}
\times \biggl \{\pi Q^2 {{\hat t}\over{\hat u}}+{{\pi}\over {3}}
{{Q^2(Q^2\hat u-\hat s \hat t)^2} \over{\hat u \hat t (\hat
s-Q^2)^2}}+{{4\pi}\over{3}} {{Q^4\hat s}\over{(\hat s-Q^2)^2}}
\biggr \},
\end{equation}
$|M(+-;-++)|^2=|M(+-;+-+)|^2$\,\,\, for \,\,\,
$R_e\longleftrightarrow L_e$,
$$
|M(+-;+--)|^2= \biggl[\sum_{n=1}^{n_z}2|D_{Z_n}(q^2)|^2\,g_s^2 g^4
L_{q_n}^2 L_{e_n}^2+\frac{8g_s^2 e^4 e_q^2}{q^4}+
$$
$$
+\sum_{\scriptstyle n,n'=1 \atop\scriptstyle n<n'}^{n_z} 4 \cdot
Re[D_{Z_n}(q^2)D_{Z_{n'}}^\star(q^2)] g_s^2 g^4 L_{q_n}L_{q_{n'}}
L_{e_n}L_{e_{n'}}+
$$
$$
+\sum_{n=1}^{n_Z}\frac{[|D_{Z_n}(q^2)|^2 8 g_s^2 g^2 e_q L_{q_n}
L_{e_n}(q^2 - m_{Z_n}^2)]}{q^2}\biggr] \times
$$
$$
\times \biggl \{ {{2\pi}\over{3}} {{(Q^2 \hat t -\hat s \hat
u)^2}\over{\hat u (\hat s - Q^2)^2}}
+{{\pi}\over {3}} {{\hat s (Q^2 \hat t -\hat u \hat s)^2} \over{\hat u
\hat t (\hat s-Q^2)^2}}
+{{4\pi}\over{3}} {{Q^2 \hat s \hat t} \over{(\hat s-Q^2)^2}} +
$$
\begin{equation}
 +{{4\pi}\over{3}} {{Q^2 \hat s^2 } \over{(\hat s-Q^2)^2}} -
2\pi \hat u - 2\pi Q^2 -\pi {{\hat s \hat u }\over {\hat t}} - 2
\pi Q^2 {{\hat s}\over{\hat t}} \biggr \},
\end{equation}
$|M(+-;-+-)|^2=|M(+-;+--)|^2$\,\,\, for \,\,\, $R_e \longleftrightarrow
L_e$. \\
The following abbreviation have been used
$$D_{Z_n}(q^2)=\frac{1}{q^2 - m_{Z_n}^2+im_{Z_n} \Gamma_{Z_n}}.$$
For $n_Z=1$ one recovers the cross section of the MSSM [10]. In
models $R5_1$ and $R5_2$ the number of neutral gauge bosons is
$n_Z$=2, in models $R6$ $n_Z$=3. Note that all couplings are
assumed to be real due to CP conservation.

The experimental lower mass bounds on the new $E_6$ gauge bosons
are about 600\,\,GeV [14]. For calculation we assume $m_{Z_2}$ =
1264\,\,GeV in the model $R5_1$, $m_{Z_2}$= 1786\,\,GeV in $R5_2$
and $m_{Z_3}$= 1786\,\,GeV in $R6$. The widths of the new gauge
bosons are estimated by $\Gamma_{Z_{2,3}}$=0.014$\,\, m_{Z_{2,3}}$
[15]. All scalar products $p_i \cdot  p _j$  can be expressed in
terms
of Mandelstam variables [13]: \\
$$\hspace*{23mm} s_{12} =  \hat s, $$
$$\hspace*{15mm} s_{13}=2p_1p_3={1\over 2}(Q^2 - \hat t) - {{Q^2 \hat u
- \hat s \hat t}
\over{2( \hat s-Q^2)}} \cos \alpha -{{\sqrt{Q^2 \hat s \hat t \hat
u}}\over{\hat s-Q^2}}\sin\alpha \cos\beta,$$
$$\hspace*{15mm} s_{14}=2p_1p_4={1\over 2}(Q^2-\hat t) + {{Q^2\hat
u-\hat s \hat t}
\over{2(\hat s-Q^2)}}\cos\alpha +{{\sqrt{Q^2\hat s \hat t\hat
u}}\over{\hat s -Q^2}}\sin\alpha \cos\beta,$$
$$\hspace *{18mm} s_{15} =2 p_1 p_5 =-\hat u $$
\begin{equation}
s_{23}=2p_2p_3={1\over 2}(Q^2-\hat u) - {{Q^2 \hat t- \hat s \hat u}
\over{2(\hat s-Q^2)}}\cos\alpha +{{\sqrt{Q^2 \hat s\hat t\hat u}}\over{
\hat s-Q^2}}\sin\alpha \cos\beta,
\end{equation}
$$\hspace*{15mm} s_{24}=2p_2p_4={1\over 2}(Q^2- \hat u) + {{Q^2 \hat
t-\hat s\hat u}
\over{2(\hat s-Q^2)}}\cos\alpha -{{\sqrt{Q^2 \hat s \hat t \hat
u}}\over{ \hat s-Q^2}}\sin\alpha \cos\beta,$$
$$\hspace*{20mm} s_{25}=2p_2p_5=-\hat t, $$
$$\hspace*{20mm} s_{34}=2p_3p_4=Q^2,$$
$$\hspace*{20 mm}s_{35}=2p_3p_5=-{{\hat u+ \hat
t}\over{2}}(1-\cos\alpha),  $$
$$\hspace*{20 mm}s_{45}=2p_4p_5=-{{\hat u+ \hat t}\over{2}}(1+\cos
\alpha) . $$ \\
Integration over the final states in phase space can be simplified by
employing the relation
\begin{equation}
{{1}\over{(2\pi)^9}}{{d^3p_3}\over{2E_3}}{{d^3p_4}\over{2E_4}}
{{d^3p_5}\over{2E_5}} \delta (p_1+p_2-q-p_5) ={{1}\over{(2\pi)^9}}
{{1}\over{16}} d\Omega \pi \delta (\hat s+ \hat t+ \hat u-Q^2){{dQ^2d
\hat td \hat u}\over{\hat s}}.
\end{equation}
The effective cross section for $pp \to l^+l^- +X$ processes can be
represented in the form [16]
\begin{equation}
E{{d \sigma}\over{dQ^2d^3p}}= \int\limits_{x_1^{min}}^1
\int\limits_{x_2^{min}}^1 dx_1 dx_2 G^A(x_1)G^B(x_2) {{\hat
s}\over{\pi}} {{d\hat \sigma}\over{dQ^2d \hat t d \hat
u}}\delta(\hat s+ \hat t+\hat u-Q^2),
\end{equation}
$$\pi E {{d\sigma}\over{d^3p}}={{d\sigma}\over{dydp_T^2}}.$$
where $y$ is the rapidity of the lepton pair, $p_T$  is its
transverse momentum, and $G^A(x_1)$ and $G^B(x_2)$  are distributions of
the partons in the proton. From expression (21), it follows that, in the
double-spin case, the correlation effective cross section has a form
\begin{equation}
{{d\Delta \sigma}\over{dQ^2 dy dp_T^2}}= \int\limits_{x_1^{min}}^1
\int\limits_{x_2^{min}}^1 dx_1 dx_2 \Delta G^A(x_1) \Delta
G^B(x_2) \hat s {{d \Delta \hat \sigma}\over{dQ^2d\hat t  d\hat
u}}\delta(\hat s + \hat t+\hat u-Q^2),
\end{equation}
$$d\Delta \sigma={1\over2} (d\sigma^{(++)} -d\sigma^{(+-)}), $$
whereas, in the single-spin case, we obtain
\begin{equation}
{{d \Delta \sigma}\over{dQ^2dydp_T^2}}=
\int\limits_{x_1^{min}}^1 \int\limits_{x_2^{min}}^1 dx_1 dx_2 \Delta
G^A(x_1)
G^B(x_2) \hat s {{d \Delta \hat \sigma}\over{dQ^2d\hat td \hat u}}
\delta(\hat s+\hat t+\hat u-Q^2),
\end{equation}
$$
d \Delta \sigma = d \sigma^{(+)} - d \sigma^{(-)} = {1 \over 2}(d
\sigma^{(++)}
+d \sigma^{(+-)} -d \sigma^{(-+)} - d \sigma^{(--)}),
$$
$$\hat s=x_1x_2s,$$
$$\hat t=x_1t+(1-x_1)Q^2,$$
\begin{equation}
\hat u=x_2u+(1-x_2)Q^2,
\end{equation}
$$t=Q^2 -m_T \sqrt{s} e^{-y},$$
$$u=Q^2 -m_T \sqrt {s} e^y $$
where $Q^2$ is the invariant mass of the lepton pair and $m_T$ -- is
the transverse mass, which is given by
$$m_T^2=Q^2+p_T^2;$$
\begin{equation}
x_1={{x_2 \sqrt {s} \sqrt {Q^2+p_T^2} e^y-Q^2}\over{x_2s- \sqrt {s}
\sqrt {Q^2+p_T^2} e^{-y}}};  \quad
x_2= {{x_1 \sqrt {s}  \sqrt {Q^2+p_T^2} e^{-y}-Q^2}\over{x_1s- \sqrt
{s} \sqrt {Q^2+p_T^2} e^{y}}};
\end{equation}
$$x_1^{min}= {{-u}\over{s+t-Q^2}}={{\sqrt {s}
\sqrt{Q^2+p_T^2}e^y-Q^2}\over{s- \sqrt {s} \sqrt{Q^2+p_T^2} e^{-y}}};$$
\begin{equation}
x_2^{min}= {{-t}\over{s+t-Q^2}}={{\sqrt {s}
\sqrt{Q^2+p_T^2}e^{-y}-Q^2}\over{s- \sqrt {s} \sqrt{Q^2+p_T^2} e^y}}.
\end{equation}
In order to compute single- and the double-spin asymmetries, we
introduce the quantities
$$
d\hat \sigma^{(++)} \pm d \hat \sigma^{(+-)} \sim \biggl \{
(|M(++;+-+)|^2+|M(++;-++)|^2 \pm |M(+-;+--)|^2 \pm
$$
$$
\pm |M(+-;-+-)|^2) \pm(|M(+-;+-+)|^2 + |M(+-;-++)|^2 +
$$
\begin{equation} + |M(+-;+--)|^2 + |M(+-;-+-)|^2 \biggr \},
\end{equation}
$$
d\hat \sigma^{(+)} \pm d\hat \sigma^{(-)} \sim \biggl \{
(|M(++;+-+)|^2+|M(++;-++)|^2 + |M(+-;+--)|^2  +
$$
$$
+|M(+-;-+-)|^2  \pm |M(-+;+-+)|^2 \pm |M(-+;-++)|^2 \pm |M(--;+--)|^2
\pm
$$
$$
\pm|M(--;-+-)|^2) + (|M(+-;+-+)|^2 + |M(+-;+--)|^2 +
$$
$$
+ |M(+-;-++)|^2+|M(+-;-+-)|^2 \pm |M(-+;+--)|^2 \pm
$$
\begin{equation}
\pm |M(-+;+-+)|^2 \pm |M(-+;-++)|^2 \pm |M(-+;-+-)|^2)\biggr\},
\end{equation}
For this purpose, we also use the well-known expressions
\begin{equation}
A_L = {{{{d\sigma^{(+)}} \over{dQ^2 dydp_T^2}}
-{{d\sigma^{(-)}}\over{dQ^2dydp_T^2}} } \over {
{{d\sigma^{(+)}}\over{dQ^2dydp_T^2}} +
{{d\sigma^{(-)}}\over{dQ^2dydp_T^2}}} },
\end{equation}
\begin{equation}
A_{LL}= {{{ {d \sigma^{(++)}} \over{dQ^2 dydp_T^2}}
-{{d\sigma^{(+-)}}\over{dQ^2dydp_T^2}}} \over
{{{d\sigma^{(++)}}\over{dQ^2dydp_T^2}}+{{d\sigma^{(+-)}}\over{dQ^2dydp_T^2}}}}.
\end{equation}

\section{\bf Numerical results and discussion}

In order to compute single- and the double-spin asymmetries
numerically, we employ two functions that describe the
distribution of polarized quarks which were proposed by Cheng et
al. [17] (the CLW model) and by Gluck et al. [18] (the GRSV
model). We use two sets of parameters for each function. For
unpolarized quarks the distribution function was found by Martin
et al.[19].

In this paper  we have studed the dependences of the single-$A_L$ and the double-spin $A_{LL}$
asymmetries on the dilepton transverse momentum at RHIC energies ($\sqrt s=500$ GeV) for various
values of the dilepton invariant mass: $Q=10$ GeV, $60$ GeV and $Q=m_{Z^{0}}$.

The single-spin asymmetry in $pp \to l^+l^- +X$ processes as a
function of the transverse momentum of the lepton pair is shown in
Figs.2-4 at $\sqrt s =500$ GeV for three values of the dilepton
invariant mass. The single-spin asymmetry $A_L$ as a function of
the dilepton transverse momentum $p_T$ is presented Fig.2  at the
dilepton invariant mass of $Q=10$ GeV and the rapidity of $y=0$.
As it is seen for the single-spin asymmetry $A_L$, the result
within the CLW model differs a little from that within the GRSV
model. Also, as it is seen from Figs.2-4 the dependence of the
single-spin asymmetry $A_{L}$ on the dilepton transverse momentum
demonstrates the same behavior. Single-spin asymmetry is
monotonuosly increase with increase of the dilepton transverse
momentum. As the dilepton invariant mass varies from $Q=10$ GeV to
$Q=m_{Z^{0}}$, the single-spin asymmetry can take either positive
or negative values. At the dilepton invariant mass $Q=10$ GeV for
the single-spin asymmetry $A_{L}$, the result within the CLW model
differs from that within the GRSV model and is equal to 2.4
percent. Also, at the dilepton invariant mass $Q=60$GeV and
$m_{Z^{0}}$ for single-spin asymmetries differs between CLW and
GRSV model approximately 3 and 4 percent. It should be noted that,
over the entire invariant-mass range under consideration, the
single-spin asymmetries in the GRSV model are almost independent
of the choice of the set of partons' distribution amplitudes.

The double-spin asymmetry in $pp \to l^+l^- + X$ processes as a
function of the dilepton transverse momentum at the rapidity of
$y=0$ and the energy of $\sqrt s= 500 $ GeV is illustrated in
Figs.5-7 for three values of the dilepton invariant mass: $Q=10$
GeV, $60$ GeV, $m_{Z^0}$ and the rapidity of $y=0$. For all values
of the dilepton invariant mass, the double-spin asymmetry is
greater than the single-spin asymmetry for $p_T$ values considered
in these figures. As it is seen from Figs.5-7 the dependence of
the double-spin asymmetry $A_{LL}$ on the dilepton transverse
momentum demonstrates the same behavior. Double-spin asymmetry is
monotonuosly decreasing with increase of the dilepton transverse
momentum. At the dilepton invariant mass $Q=10$ GeV, $60$ GeV and
$m_{Z^{0}}$ for double-spin asymmetries differs between CLW and
GRSV dodels approximately 7; 5 and 4 percent, respectively.

The Figs.8-9 show contribution of $Z'$ boson into the single-and the
double -spin asymmetries as the function of the dilepton
transverse momentum at the value of the dilepton invariant mass $Q
= m_{Z^{0}}$(single-spin asymmetry), 10 GeV (double-spin asymmetry).

In the present work we performed numerical analysis of single-and
double- spin asymmetries for $ pp \to (\gamma^*,Z^{0},Z') + X $
processes using the method of helicity amplitudes. Asymmetries
were explored in the domain of high momentum transfers at RHIC
energies. We employed the distributions of polarized partons
within  CLW [17] and GRSV [18] models. Both the CLW and the GRSV
functions were obtained in the second order of perturbation
theory. The distribution of unpolarized partons was taken from
[19]. In general, the distinction between the single-spin
asymmetries $A_L$ for the two sets of partons in CLW model is
greater than that in GRSV model. These asymmetries generally
increase as the dilepton mass approaches the $Z^{0}$ -boson mass.
The distinction between the two sets in GRSV model is small.
However, as the transverse momentum $p_T$,  varies within the
range $\sim 10 \div 110$ Gev/c, the double-spin asymmetries
$A_{LL}$ are of the same order of magnitude at the three values of
$Q$. At various values of $p_T$, the difference between the
asymmetries for the two sets of partons in CLW model is greater
than that in GRSV model. The difference of the asymmetries for the
two sets in GRSV model is nearly constant.

The contribution of the $Z^{0}$ and $Z^{'}$- bosons processes into polarized
particles, in general, and to the single-spin parity-violating
asymmetries, in particular, is significant. The single-spin
asymmetries become large, both in CLW and in GRSV models,
as the dilepton invariant mass approaches the $Z^{0}$ - boson
mass. Over the range of $p_T$ under study, the double-spin
asymmetry is greater than the single-spin asymmetry for all values
of the dilepton mass.

Analysis shows that with increase of mass of superstring $Z'$ boson
character dependence asymmetries on the tranverse momentum of
lepton pair  is not changed. Also, we compared this calculation
with [10]. Analysis shows that at the value of dilepton invariant
masses $Q=10$\,\,GeV, $6$0\,\,GeV, $m_{Z^{0}}$ contribution coming
from $Z'$ boson on the single-and the double -spin asymmetries is
considerable. Contribution of superstring $Z'$ boson is evident in the
interference term. Also pure contribution coming from superstring $Z'$ boson
into the single-and the double-spin asymmetries has been extracted. It is shown
that with increase of mass  of this boson to the single-and the double-spin asymmetries increase.
In general, we can show that single-and the double-spin
asymmetries have the form:
$$A_{L(LL)}=A_{L(LL)}^{\gamma,Z^{0}} + \Delta A_{L(LL)}$$
where  $\Delta A_{L(LL)}$ consist of interference term between
superstring $Z'$  boson with $Z^{0}$ and ${\gamma}$ boson and
contribution coming from superstring $Z'$ boson, with lepton pair
production from $Z'$ boson. Our calculation shows that
contribution of $\Delta A_{L(LL)}$ into single-spin asymmetries at
the value of dilepton invariant masses $Q= 10\, GeV, 60\, GeV,
m_{Z^{0}}$ is about $7.1\div 10$, $16.9 \div 19.9$, $19.4 \div
23.3$ and to the double -spin asymmetries is about $23.3 \div
28.8$, $22.8 \div 26.8$, $16.9 \div 20.9$ percent depending on
transverse momentum of the lepton pair respectively.
Both the single-spin and the double-spin asymmetries are sensitive to
the polarised gluon distribution and may be used as probes of the spin
structure of the proton

\section{\bf Conclusions}

In this paper we have studied both the single-and the double-spin
asymmetries of the transverse momentum lepton pairs at RHIC
energies ($\sqrt s$ = 500 GeV). At these high energies, $Z^{0}$
can contribute significantly into the Drell-Yan process, inducing a
parity -violating single-spin asymmetry. This asymmetry may be
measured in experiments where only one of the initial particles is
polarized. Using these angular distributions, we have constructed
single-and double-spin asymmetries, which we then studied
numerically using four sets, which both sets (CLW and GRSV) were
obtained in the second order of perturbation theory. We  have found
that the single-spin asymmetries are measurably large only at large
dilepton masses, i.e.close to the $Z^{0}$ peak. The double-spin
asymmetries are large even at smaller dilepton masses. Both the
single-spin and double-spin asymmetries are sensitive to the
polarized gluon distribution and may be used as probes of the spin
structure of the proton. As it is shown on figures with increase  of
mass of superstring $Z^{\prime}$ boson character dependence asymmetries on
the tranverse momentum  of lepton pair  is not changed. Analysis
shows that at the value of dilepton invariant masses $Q=10$
GeV, $60$ GeV, $m_{Z^0}$ contribution coming from $Z'$ boson into
the single-and the double-spin asymmetries is considerable.
Contribution of superstring $Z'$ boson is evident in the
interference term. Also, pure contribution coming from superstring $Z'$ boson
on the single-and double-spin asymmetries has been extracted. It is shown that
with increase of mass  of this boson the single-and the double-spin asymmetries increase. Our
calculation shows that contribution of $\Delta A_{L(LL)}$ into
single-spin asymmetries at the value of dilepton invariant masses
$Q= 10\, GeV, 60\, GeV, m_{Z^{0}}$ is about $7.1\div 10$, $16.9
\div 19.9$, $19.4 \div 23.3$ and to the double -spin asymmetries
is about $23.3 \div 28.8$, $22.8 \div 26.8$, $16.9 \div 20.9$
percent depending on transverse momentum of the lepton pair
respectively.

Therefore, measurement of the single-and the double-spin asymmetry helps in
study of the spin structure of the proton.

\vspace*{1.0cm}

{\bf Acknowledgment \\}

One of us Dr. A.~Ahmadov grateful to NATO Reintegration
Grant-980779.

\newpage

\begin{figure}[htb]
  \begin{center}
\mbox{\epsfig{figure=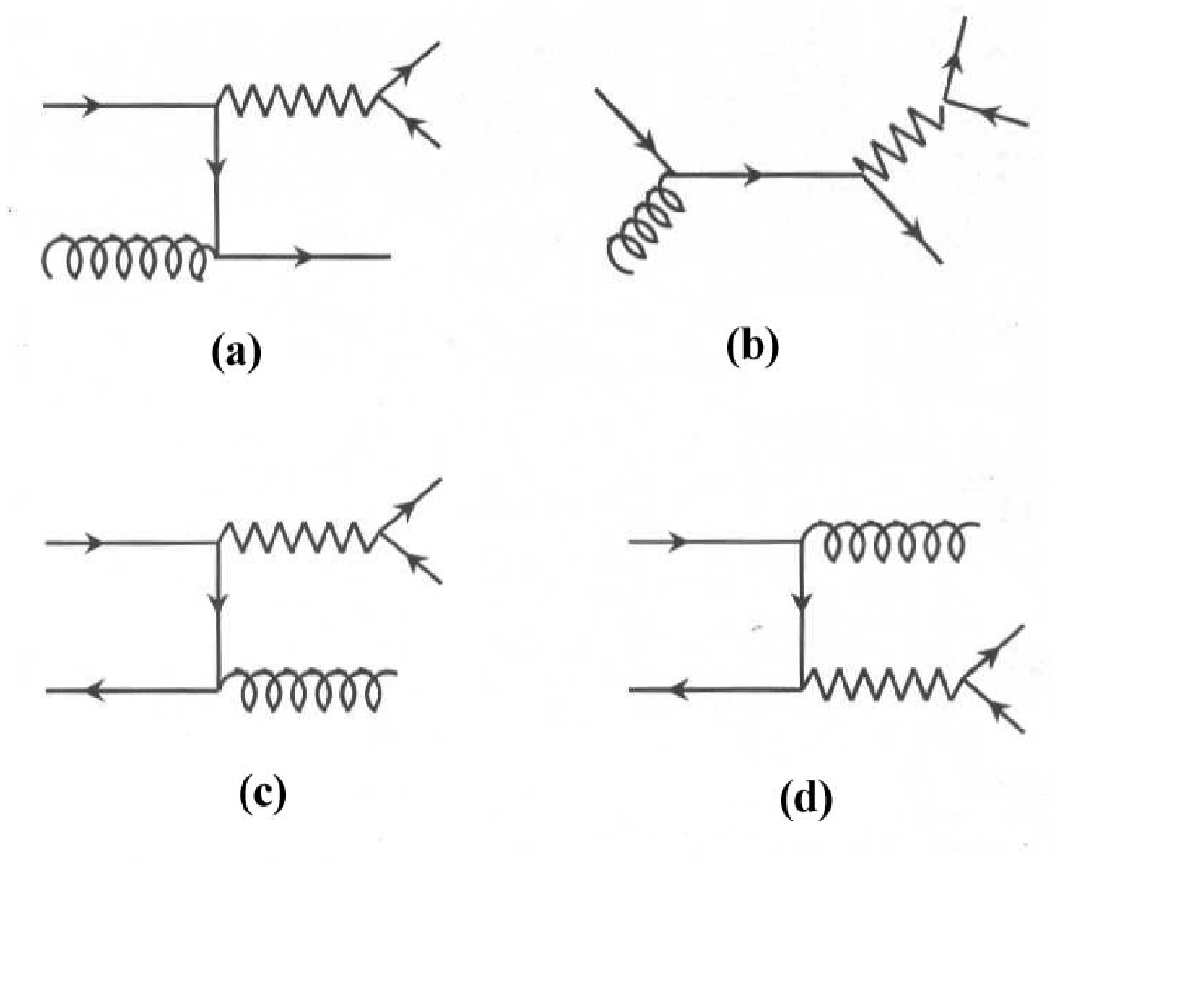,height=10cm}}
  \end{center}
\caption{Parton-level subprocesses contributing into the Drell-Yan
process: (a),(b)-quark-gluon Compton scattering;  (c),(d)-real
gluon corrections to $q \bar q$ annihilation.} \label{Fig1}
\end{figure}

\newpage

\begin{figure}[hpb]
  \begin{center}
\mbox{\epsfig{figure=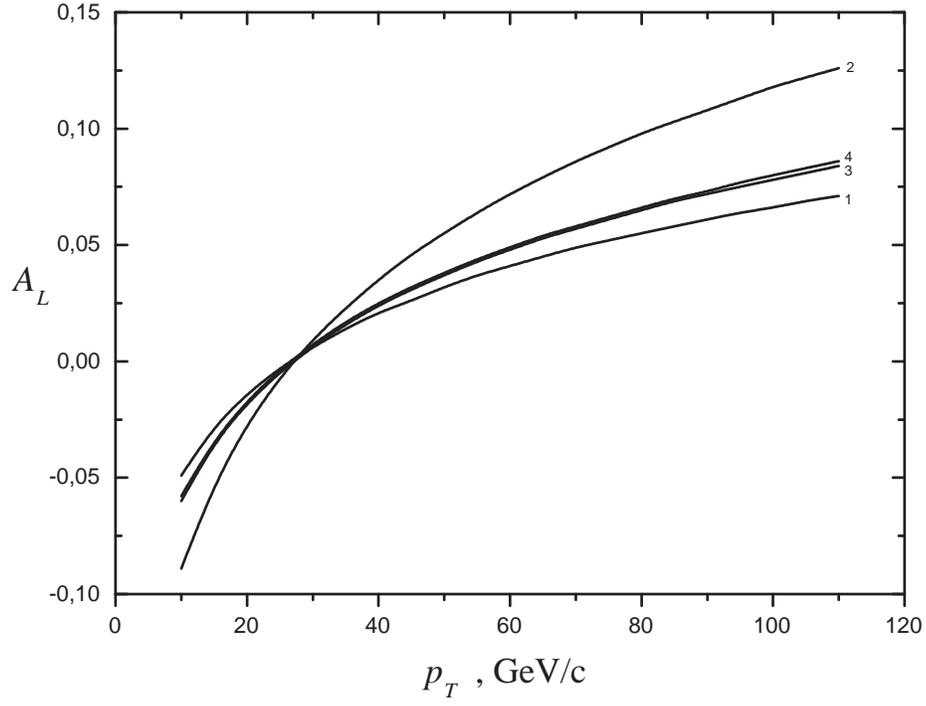,height=11cm}}
  \end{center}
\caption{Single-spin asymmetry in $pp\to l^+l^- +X$ processes as a
function of the dilepton transverse momentum at $\sqrt s=500$ GeV,
the dilepton invariant mass of $Q=10$ GeV, and the rapidity of
$y=0$. Shown in the figure are the results obtained on the basis
of the CLW model with set I (curve 1) and set II (curve 2) and on
the basis GRSV model with set I (curve 3) and set II (curve 4).}
\label{Fig2}
\end{figure}

\newpage

\begin{figure}[hpb]
  \begin{center}
\mbox{\epsfig{figure=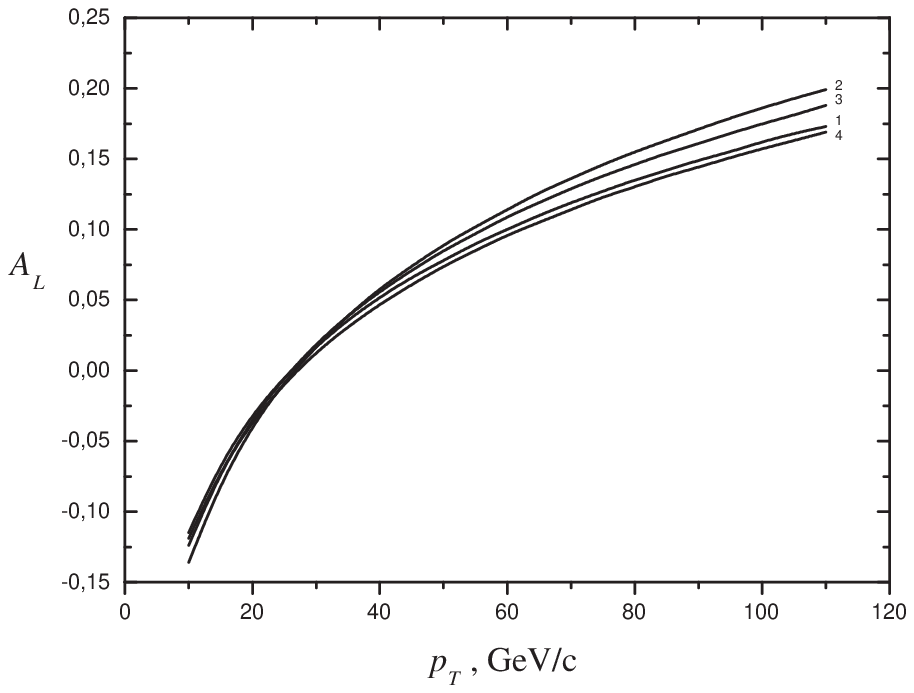,height=11cm}}
  \end{center}
\caption{As in Fig.2, but for $Q=60$ GeV.}
\label{Fig3}
\end{figure}

\newpage

\begin{figure}[hpb]
  \begin{center}
\mbox{\epsfig{figure=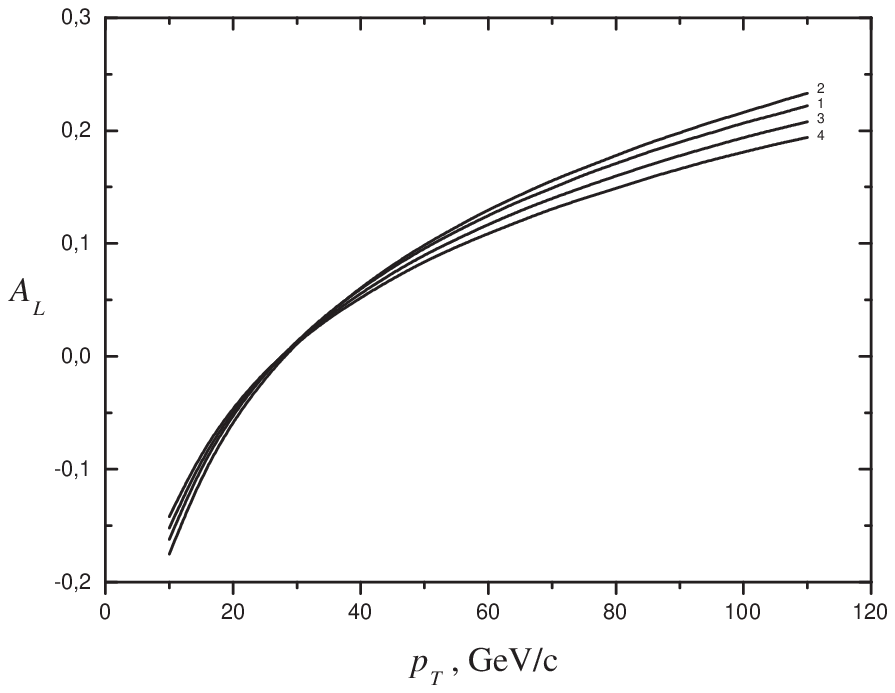,height=11cm}}
  \end{center}
\caption{As in Fig.2, but for $Q=m_{Z^0}$.}
\label{Fig4}
\end{figure}

\newpage

\begin{figure}[hpb]
  \begin{center}
\mbox{\epsfig{figure=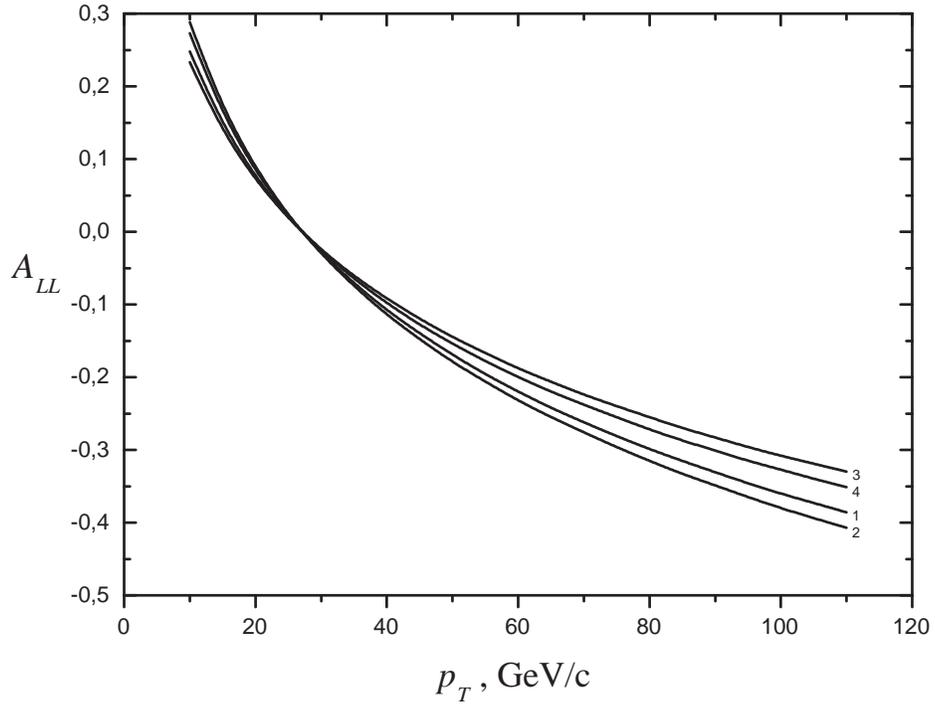,height=11cm}}
  \end{center}
\caption{Double-spin asymmetry $pp \to l^+l^- +X$ in the processes
as a function of the dilepton transverse momentum at $\sqrt s
=500$ GeV, the dilepton invariant mass of $Q=10$ GeV, and the
rapidity of $y=0$. Shown are the asymmetries obtained on the basis
of the CLW model with (curve 1) set I and (curve 2) set II and on
the basis of the GRSV model with (curve 3) set I and (curve 4) set
II. }
\label{Fig5}
\end{figure}

\newpage

\begin{figure}[hpb]
  \begin{center}
\mbox{\epsfig{figure=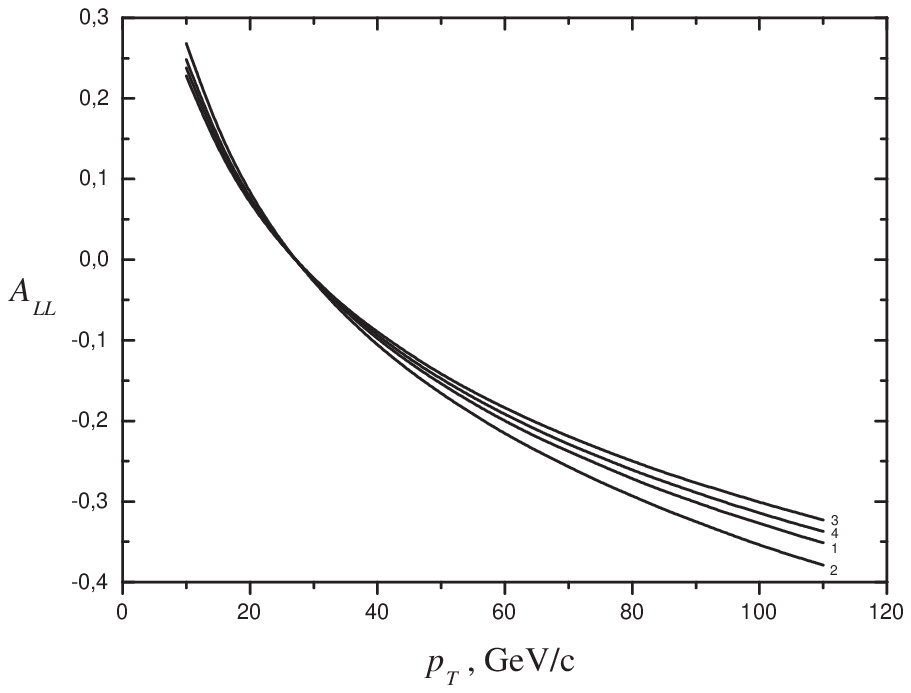,height=11cm}}
  \end{center}
\caption{As in Fig.5, but for $Q=60$ GeV.}
\label{Fig6}
\end{figure}

\newpage

\begin{figure}[hpb]
  \begin{center}
\mbox{\epsfig{figure=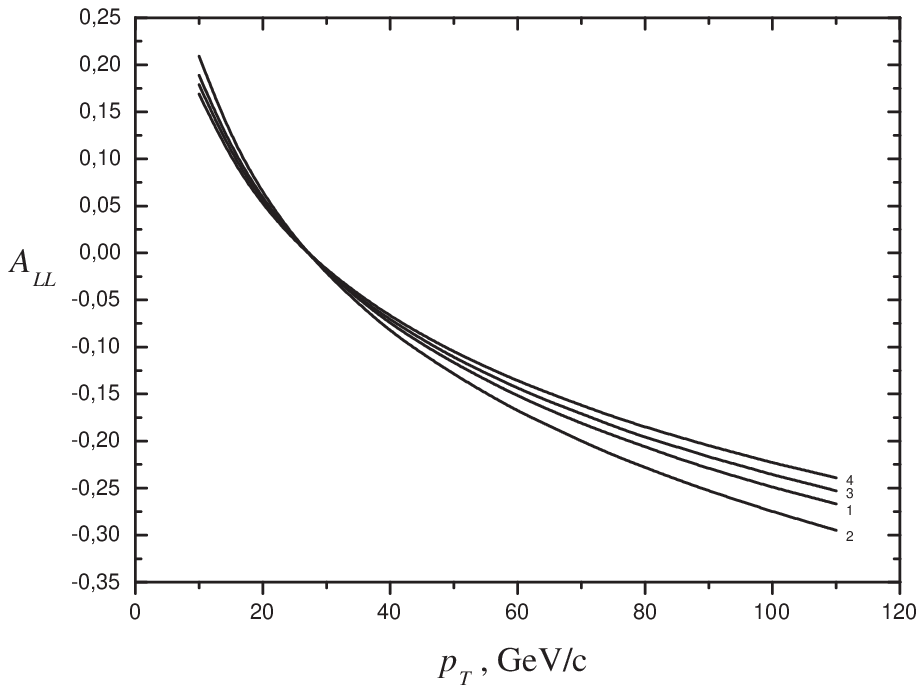,height=11cm}}
  \end{center}
\caption{As in Fig.5, but for $Q=m_{Z^0}$.}
\label{Fig7}
\end{figure}

\newpage

\begin{figure}[hpb]
  \begin{center}
\mbox{\epsfig{figure=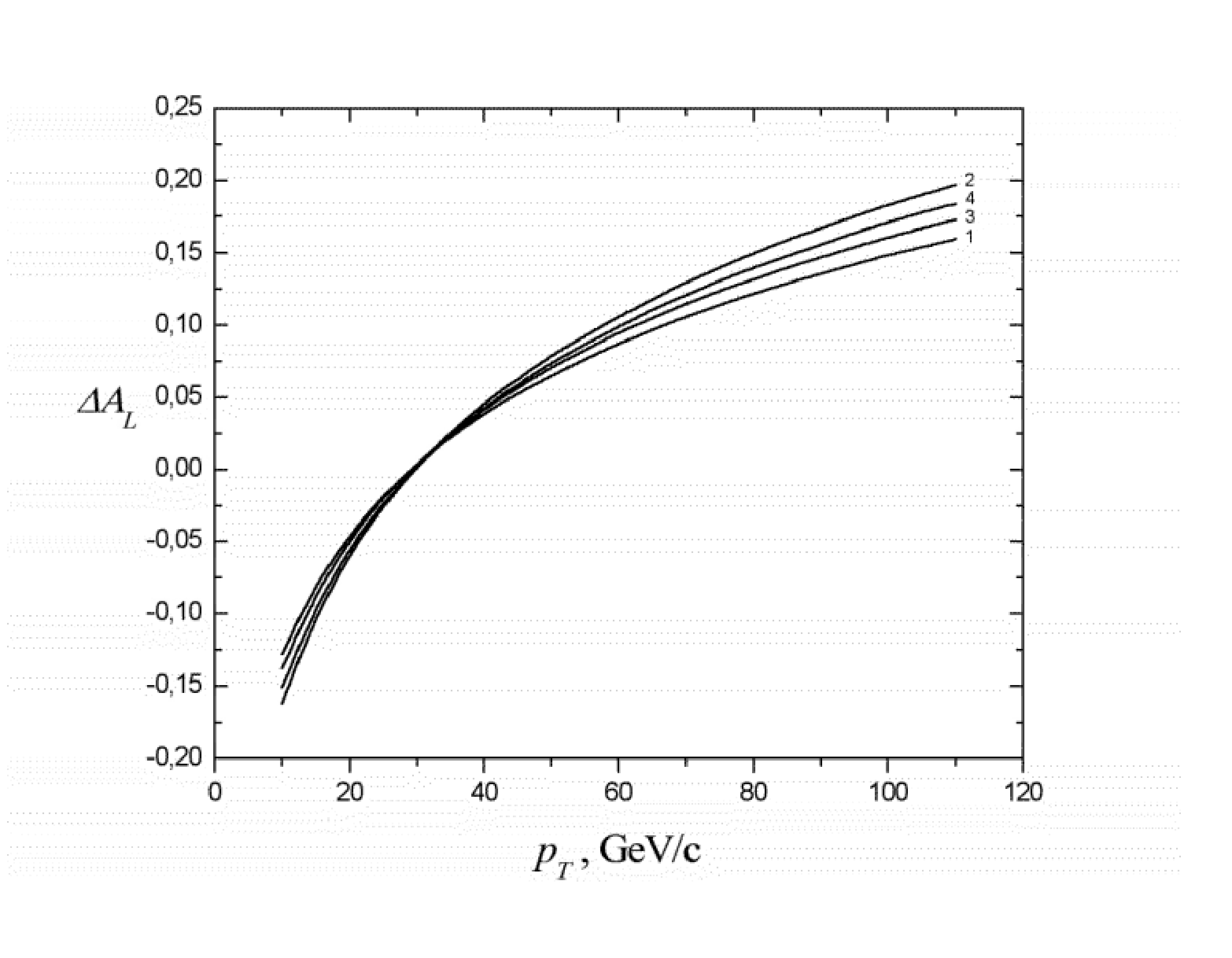,height=12cm}}
  \end{center}
\caption{The dependence contribution $Z'$ boson on the single-spin
asymmetries $\Delta A_L$ as the function of the dilepton
transverse momentum at $\sqrt s =500$ GeV, the dilepton invariant
mass of $Q=m_{Z^0}$, and the rapidity of $y=0$. Shown are the
asymmetries obtained on the basis of the CLW model with set I
(curve 1) and set II (curve 2) and on the GRSW model with set I
(curve 3) and set II (curve 4).}
\label{Fig8}
\end{figure}

\newpage

\begin{figure}[hpb]
  \begin{center}
\mbox{\epsfig{figure=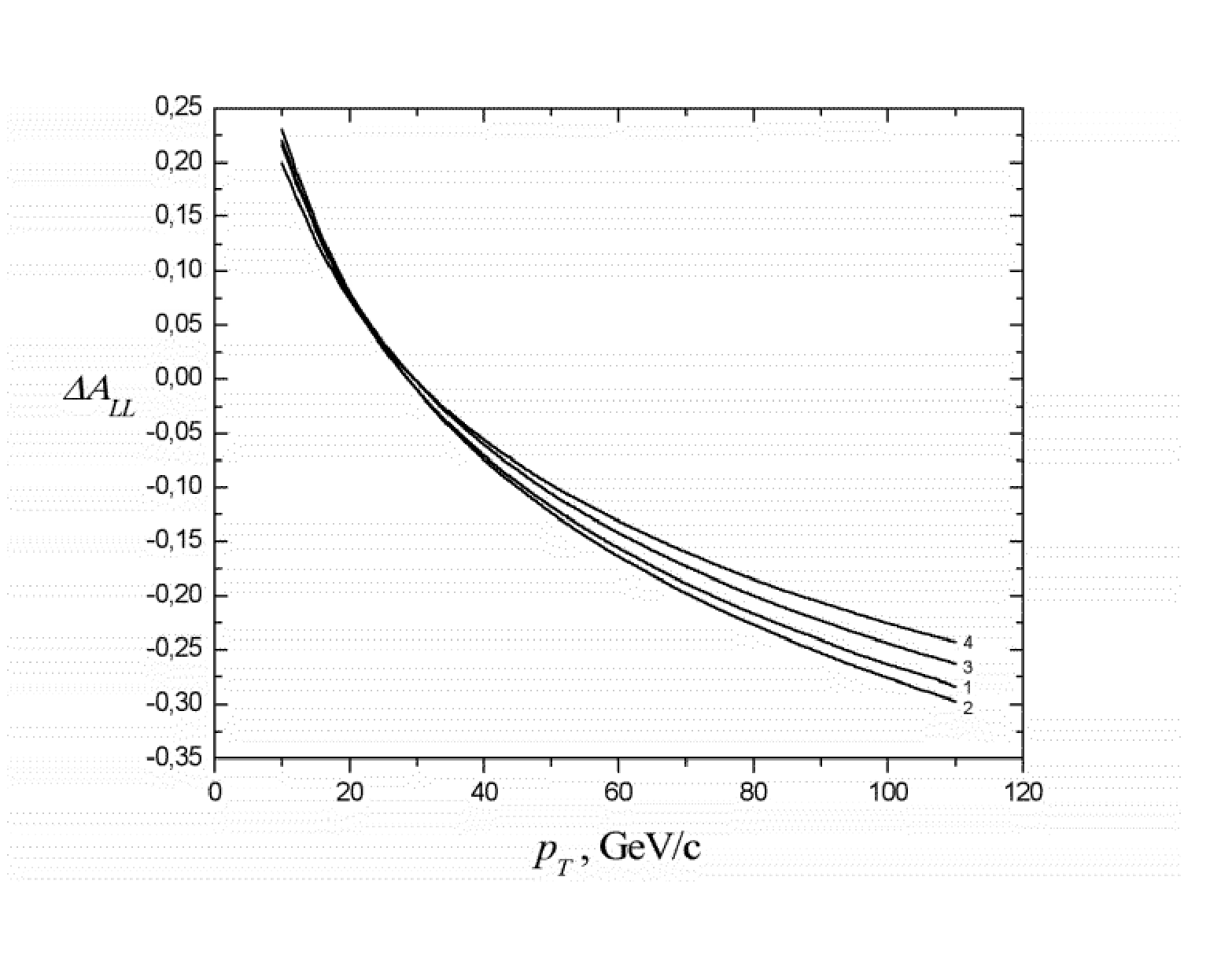,height=12cm}}
  \end{center}
\caption{The dependence contribution $Z'$ boson on the double-spin
asymmetries $\Delta A_{LL}$ as the function of the dilepton
transverse momentum at $\sqrt s=500$ GeV, the dilepton invariant
mass of $Q=10$ GeV, and the rapidity of $y=0$. Shown are the
asymmetries obtained on the basis of the CLW model with set I
(curve 1) and set II (curve 2) and on the GRSW model with set I
(curve 3) and set II (curve 4).} \label{Fig9}
\end{figure}

\end{document}